\documentclass[aps,prl,twocolumn,amsmath,amssymb,superscriptaddress,floatfix,superscriptaddress,longbibliography]{revtex4-2}

\usepackage{graphicx}
\usepackage{multirow}
\usepackage[dvipsnames]{xcolor}
\usepackage{bm}
\usepackage{amsfonts,amssymb,amsmath}
\usepackage{graphicx,dcolumn,bm,color,braket,slashed}
\usepackage{times} 
\usepackage{comment, ulem}

\def\ket#1{|#1\rangle }

\begin{document}

\title{Optical Signatures and Quantum Geometry in Proximity-Induced Topological Superconductors}

\author{Myungjun Kang}
\thanks{These authors contributed equally}
\affiliation{Department of Physics, Hanyang University, Seoul 04763, Korea}
\affiliation{High Pressure Research Center, Hanyang University, Seoul 04763, Korea}
\affiliation{Department of Physics, University of Wisconsin--Milwaukee, Milwaukee, Wisconsin 53201, USA}

\author{Yogeshwar Prasad}
\thanks{These authors contributed equally}
\affiliation{Department of Physics, Hanyang University, Seoul 04763, Korea}
\affiliation{Research Institute of Basic Sciences, Seoul National University, Seoul 08826, Korea}

\author{Nikhil Danny Babu}
\thanks{These authors contributed equally}
\affiliation{Department of Physics, Hanyang University, Seoul 04763, Korea}

\author{Rasoul Ghadimi}
\affiliation{Department of Physics, Hanyang University, Seoul 04763, Korea}
\affiliation{Research Institute of Basic Sciences, Seoul National University, Seoul 08826, Korea}

\author{Jae Hoon Kim}
\affiliation{Department of Physics, Yonsei University, Seoul 03722, Korea}

\author{Sangmo Cheon}
\email{sangmocheon@hanyang.ac.kr}
\affiliation{Department of Physics, Hanyang University, Seoul 04763, Korea}
\affiliation{High Pressure Research Center, Hanyang University, Seoul 04763, Korea}
\affiliation{Research Institute for Natural Science, Hanyang University, Seoul 04763, Korea}
\affiliation{Hanyang Institute for Quantum Science and Quantum Technology, Hanyang University, Seoul 04763, Korea}

\begin{abstract}
Topological-insulator-superconductor (TI-SC) heterostructures provide a promising platform for proximity-induced topological superconductivity, but diagnosing superconductivity at a buried interface remains challenging for conventional surface-sensitive probes. Here, we develop a quantitative theory of the longitudinal optical response of a TI-SC heterostructure and show that the complex sheet conductance provides an interface-selective route to isolating and diagnosing the buried proximitized interface state. Starting from a minimal model, we derive a low-energy description of the heterointerface in which the induced gap emerges directly from the TI-SC coupling. Combined with a slab-based thickness-extrapolation procedure, this framework yields a practical protocol for separating the buried interfacial sheet conductance from bulk and exposed-surface optical contributions. The extracted interface response exhibits a robust, thickness-independent coherence peak at an energy set by the proximity-induced gap, clearly distinguishable from both the pair-breaking scale of the parent superconductor and the Dirac cone on the exposed TI surface.
At low energies, the heterointerface is described by an effective time-reversal-invariant topological-superconducting theory, while the associated low-frequency optical spectral weight admits a quantum-geometric interpretation through the optical sum rule.
Our results establish terahertz/infrared spectroscopy of thickness-extracted sheet conductance as a noninvasive route to identifying and quantifying proximity-induced superconductivity at buried TI-SC interfaces.
\end{abstract}

\maketitle

\clearpage
\section*{Introduction}

Topological superconductors (TSCs) support Majorana boundary excitations with non-Abelian exchange statistics and are therefore of broad interest as a platform for fault-tolerant quantum information processing~\cite{qi2011topological,PhysRevLett.105.097001,sarma2015majorana,hastings2013metaplectic,sanno2022engineering,you2019building,nahum2020entanglement}. 
A natural route to such phases is intrinsic odd-parity, including candidate $p$-wave, superconductivity~\cite{kitaev2001unpaired}.
However, although candidate materials such as Cu$_x$Bi$_2$Se$_3$ have been proposed~\cite{hor2010superconductivity,sasaki2011topological}, intrinsic odd-parity superconductivity is rare and is generally much more sensitive to nonmagnetic disorder than conventional isotropic $s$-wave superconductivity, for which Anderson's theorem applies~\cite{li2015local,Nagai2014,Mackenzie2003,ANDERSON195926,abrikosov1961problem}.

An alternative strategy is to engineer topological superconductivity at the interface between a three-dimensional topological insulator (TI) and a conventional $s$-wave Bardeen-Cooper-Schrieffer (BCS) superconductor through the proximity effect~\cite{fu2008superconducting,hasan2010colloquium,stanescu2010proximity,black2013proximity,trang2020conversion,park2020proximity,Potter2011}. 
In such TI-SC heterostructures, pairing induced in the helical Dirac surface states can generate an effective two-dimensional topological superconducting phase. 
Proximity-induced superconductivity has been reported in several related platforms, including nanowire-based systems~\cite{mourik2012signatures,das2012zero,deng2016majorana}, Bi$_2$Se$_3$ thin films on NbSe$_2$~\cite{wang2012coexistence}, and Bi$_2$Se$_3$/Bi$_2$Te$_3$ heterostructures coupled to cuprates~\cite{zareapour2012proximity}. 
Josephson transport, anomalous current-phase relations, Fraunhofer-pattern anomalies, and vortex-core zero-bias states have likewise been interpreted as signatures of proximity-induced topological superconductivity in TI-based platforms~\cite{williams2012unconventional,veldhorst2012josephson,xu2015experimental}. 
Yet an unambiguous diagnosis of the buried proximitized interface itself remains challenging.

\begin{figure*}[t]
\includegraphics[width=0.95\textwidth]{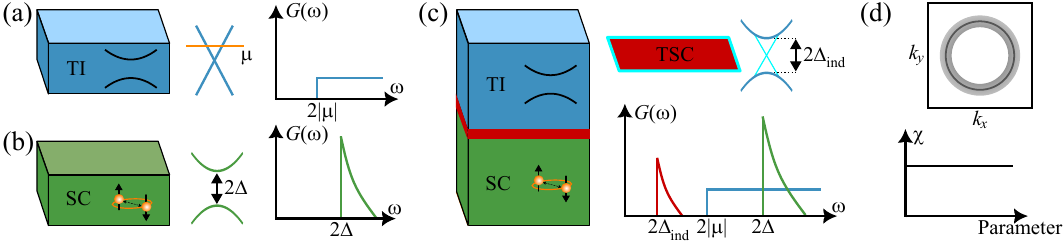}
\caption{\label{fig1:Schematics}
    \textbf{Schematics of the TI-SC heterostructure and its optical and quantum-geometric signatures.}
    \textbf{(a)} Isolated topological insulator (TI) with a gapped bulk and a gapless Dirac surface state.
    The chemical potential $\mu$ (orange) intersects the Dirac cone, producing a Pauli-blocked, step-like in-plane optical conductance $G_{xx}(\omega)$ with onset at $2|\mu|$.
    \textbf{(b)} Isolated conventional $s$-wave BCS superconductor (SC) with pairing amplitude $\Delta$.
    Its optical response is gapped below $2\Delta$ and exhibits a coherence peak at the gap edge.
    \textbf{(c)} TI-SC heterostructure obtained by placing the TI (blue) on top of the SC (green).
    At low energies, the interface region (red) is described by an effective two-dimensional time-reversal-invariant topological-superconducting theory.
    Proximity pairing opens an induced gap $2\Delta_{\text{ind}}$ in the buried TI interface Dirac cone.
    The resulting heterointerface conductance $G_{xx}(\omega)$ exhibits a coherence peak at $2\Delta_{\text{ind}}$, a second feature at $2\Delta$ from the parent SC, and a Pauli-blocking contribution from the ungapped top-surface Dirac cone.
    \textbf{(d)} Quantum-geometric interpretation of the low-frequency interfacial optical weight.
    The gray ring in momentum space marks the region where the quantum metric $g_{xx}(\mathbf{k})$ of the interface state is strongly enhanced by the induced gap.
    The corresponding quantum weight
    $K_{\mu\mu}\propto \int_{\text{BZ}} d^2k\, g_{\mu\mu}(\mathbf{k})$
    is connected to the generalized optical weight through the sum rule
    $\int_{0}^{\infty} d\omega\, \text{Re}[G_{\mu\mu}(\omega)]/\omega \propto K_{\mu\mu}$.
    As a result, the dimensionless ratio $\chi$ between these two quantities remains approximately unchanged under variations of key parameters such as the chemical potential and the proximity-coupling strength.
}
\end{figure*}

Optical and terahertz spectroscopy provide a noninvasive and contactless route to this problem because, in thin-film geometries, the relevant observable is the complex sheet conductance,
\begin{equation}
G(\omega)=\int dz\, \sigma(\omega,z),
\end{equation}
which naturally combines bulk and interface electrodynamics into a measurable quantity.
In standalone three-dimensional TIs, terahertz and optical studies of Bi-based compounds have revealed surface-dominated low-frequency transport together with nontrivial in-plane optical conductivity arising from the interplay between bulk carriers and topological surface states~\cite{valdes2012terahertz,di2012optical,tang2013terahertz,Carbotte2013a,Carbotte2013b,wu2013sudden,park2015terahertz,aguilar2015}.
By contrast, the frequency-dependent optical response of TI-SC heterostructures, especially the electrodynamics of the buried interface, remains comparatively unexplored. 
This raises the central question of whether the optical response of the buried proximitized TI-SC interface can be isolated from the exposed TI surface and the bulk superconducting background, thereby enabling direct access to the induced interfacial gap.

A complementary perspective comes from band quantum geometry. The quantum metric quantifies distances between Bloch or BdG eigenstates on the quantum-state manifold and has emerged as an important ingredient in the superfluid and optical response of multiband superconductors and flat-band systems~\cite{torma2022superconductivity,PhysRevLett.132.026002,10.1093/nsr/nwae334}. In particular, the interband contribution to the superfluid weight is governed by the Brillouin-zone integral of the quantum metric and can substantially affect the phase stiffness. Recent work has further shown that the quantum metric of BdG bands is quantitatively related, through optical sum rules, to the low-frequency optical spectral weight~\cite{ghosh2024probing,PhysRevResearch.7.023158,verma2025framework}. Thus, once the buried interface contribution is isolated, its optical response also admits a quantum-geometric interpretation in terms of the proximitized interfacial state.

Here, we develop a microscopic theory of the longitudinal optical response of a TI-SC heterostructure and show that, within a thickness-extraction framework, the complex heterointerface conductance $G_{xx}(\omega)$ provides an interface-selective diagnostic of buried proximity-induced superconductivity.
Starting from a minimal TI-SC model and integrating out the parent superconductor, we derive an explicit low-energy BdG description of the buried Dirac interface in which the induced pairing structure and the corresponding induced gap emerge analytically from the tunneling strength and the parent gap, rather than being introduced phenomenologically.
Combined with a slab-based thickness-extrapolation procedure, this microscopic-to-effective framework yields a practical route to isolate the buried interfacial sheet conductance from bulk and exposed-surface optical channels.
The resulting interfacial response exhibits a pronounced coherence feature at $\omega \simeq 2\Delta_{\text{ind}}$, providing a direct spectroscopic signature of the induced superconducting gap at the buried TI-SC interface.
Over the parameter range studied, this feature is tunable with chemical potential and weakens with temperature in tandem with the parent superconducting order parameter, while remaining discernible under moderate lifetime broadening and moderate interface-coupling inhomogeneity.

We consider a minimal TI model with surface Dirac cones, which produce a Pauli-blocked optical response [Fig.~\ref{fig1:Schematics}(a)], together with a conventional BCS superconductor whose optical conductance exhibits a coherence peak at the pair-breaking scale $2\Delta$ [Fig.~\ref{fig1:Schematics}(b)]. Placing the TI on top of the SC yields the TI-SC heterostructure shown in Fig.~\ref{fig1:Schematics}(c).
At low energies, the heterointerface (red sheet) is described by an effective two-dimensional time-reversal-invariant topological-superconducting theory.
We show that the conductance spectrum contains a coherence peak at $2\Delta_{\text{ind}}$, a second feature associated with the parent superconducting scale $2\Delta$, and a Pauli-blocked contribution from the ungapped Dirac cone on the exposed TI surface, as summarized schematically in Fig.~\ref{fig1:Schematics}(c). Using a microscopic slab model that resolves both surface and bulk electronic states, we evaluate the frequency-dependent sheet conductance $G_{xx}(\omega)$ and exploit its thickness dependence to separate bulk and interfacial optical channels. The calculations reveal a pronounced, thickness-independent resonance arising from the induced gap on the interface states, confirming that the low-frequency feature is an intrinsic property of the buried interface rather than of the TI or SC bulk backgrounds. In this way, our approach provides a controlled optical protocol for disentangling buried interfacial and bulk responses and for identifying the induced energy scale of proximity superconductivity.

We further show, through the optical-conductance sum rule~\cite{PhysRevResearch.7.023158,PhysRevX.14.011052}, that the low-frequency spectral weight of the isolated interfacial response admits a compact quantum-geometric interpretation. In the present microscopic setting, opening the proximity-induced gap at the buried Dirac interface produces a pronounced enhancement of the quantum metric near the interfacial gap edge, while the corresponding quantum weight remains quantitatively linked to the low-frequency integral $\int d\omega\, \text{Re}[G_{xx}(\omega)]/\omega$ over a broad range of model parameters, including the chemical potential and the proximity-coupling strength [Fig.~\ref{fig1:Schematics}(d)].

Taken together, these results establish thickness-extracted terahertz and infrared sheet conductance as a practical and noninvasive diagnostic of buried proximity-induced superconductivity in TI-SC heterostructures, with the associated low-frequency optical weight admitting a compact quantum-geometric interpretation.

\section*{Results}

\subsection*{
Microscopic Interface Theory and Proximity-Induced Gap}

\begin{figure*}[t]
\includegraphics[width=0.95\textwidth]{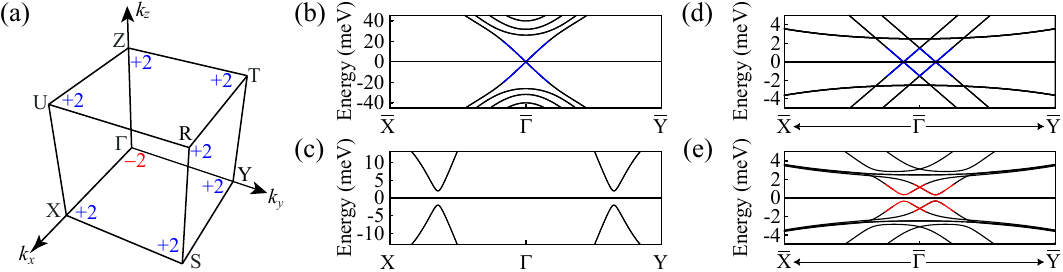}
\caption{\label{fig2:SetUp}
\textbf{Topological characteristics and band structures of the TI-SC heterostructure.}
\textbf{(a)} Parity eigenvalues of the occupied bands at the time-reversal-invariant momenta
(TRIM) for the tight-binding TI model.
A single band inversion at $\Gamma$ identifies the bulk as a TI. 
\textbf{(b)} Surface band structure of a TI slab finite along $z$ and periodic in the in-plane
directions. The resulting Dirac surface state at $\bar{\Gamma}$ is highlighted in blue.
\textbf{(c)} Bulk band structure of the $s$-wave SC, which is fully gapped and
topologically trivial. 
\textbf{(d)} Band structure of the TI-SC heterostructure near the $\bar{\Gamma}$ point for
vanishing interface hybridization $\lambda = 0$. The TI Dirac surface state (blue) remains
gapless and decoupled from the SC bands. 
\textbf{(e)} Band structure of the heterostructure for finite interface coupling $\lambda = 3$.
Hybridization with the SC induces superconducting pairing in the TI surface Dirac cone,
opening a proximity-induced gap (red) and producing a fully gapped interface spectrum
consistent with the effective BdG surface theory in Eq.~\eqref{eq:TI_SW}. 
Representative parameters are $m_0=-120$~meV, $t_0=48$~meV, $\alpha=40$~meV, $m=-45$~meV, $t=30$~meV, $\Delta=2$~meV,
$\mu=1.5$~meV, with slab thicknesses $l_T = 8$ for the TI and $l_S = 8$ for the SC.
}
\end{figure*}

Motivated by Bi$_2$Se$_3$ thin films on Nb based superconducting platforms, we model our TI-SC heterostructure using a minimal lattice Hamiltonian in which a three-dimensional TI is coupled to a conventional $s$-wave SC.
As both subsystems respect time-reversal symmetry, the TI realizes a nontrivial $\mathbb{Z}_2$ phase with a single helical Dirac surface state, while the SC provides a fully gapped order parameter.
Coupling these subsystems via a common interface induces proximity-induced pairing among the Dirac fermions.
In a slab geometry, we stack a TI layer of thickness $l_T$ on top of an SC layer of thickness $l_S$ along the $z$ direction, producing a single buried interface where the Dirac surface state hybridizes with the superconducting layer [Fig.~\ref{fig1:Schematics}(c)].
This proximitized interface defines an effective two-dimensional superconducting region.
Our goal in this section is to establish the microscopic origin of the induced interfacial gap and the corresponding low-energy theory that underlies the optical response discussed below.

The normal state of the three-dimensional TI is described by the tight-binding (TB)  Hamiltonian
\begin{eqnarray}
\label{eq:TI_Bulk}
H_{\text{TI}}(\mathbf{k})
&=&-\mu+
\left(m_0 + t_0 \sum_{i=x,y,z} \cos k_i\right)\tau_z\\
&&+\alpha \sum_{i=x,y,z} \sin k_i \,\tau_x \sigma_i,\nonumber
\end{eqnarray}
where $\tau_i$ and $\sigma_i$ are Pauli matrices acting in the orbital and spin spaces, respectively.
Here, $\mu$ denotes the chemical potential, $m_0$ controls the TI band inversion, $t_0$ the nearest-neighbor hopping amplitude, and $\alpha$ the strength of the spin-orbit coupling (SOC).
Time-reversal and inversion are represented by $T=-i\sigma_y K$ and $P=\tau_z$, respectively,
where $K$ denotes complex conjugation. 

Because the TI is centrosymmetric and time-reversal invariant, its bulk band topology is encoded in a $\mathbb{Z}_2$ index $\nu$ that can be determined from the parity eigenvalues of the occupied bands at the time-reversal-invariant momenta (TRIM) $\Gamma_i$~\cite{fu2007topological}. 
This index is given by 
\begin{equation}
(-1)^\nu
=
\prod_{i \in \text{TRIM}}
\prod_{m=1}^{N_{\text{occ}}}
\xi_{2m}(\Gamma_i) \,,
\end{equation}
where $\xi_{2m}(\Gamma_i) = \pm 1$ is the parity eigenvalue of the $2m$th occupied Kramers' pair at TRIM $\Gamma_i$, and $N_{\text{occ}}$ denotes the number of occupied Kramers' pairs. 
The individual products at each TRIM are summarized in Fig.~\ref{fig2:SetUp}(a). 
A single band inversion occurs at the $\Gamma$ point, indicating that the bulk is in a TI phase. 
For a slab geometry that is finite along $z$ but periodic in $x$ and $y$, the corresponding surface Brillouin zone is a square with surface TRIM points $\bar{\Gamma}_i$.
The band inversion at $\Gamma$ yields a single Dirac cone centered at $\bar{\Gamma}$, as highlighted by the blue surface band in Fig.~\ref{fig2:SetUp}(b).

The SC layer is modeled as a spin-degenerate single-band $s$-wave BCS superconductor with onsite pairing, described by the BdG Hamiltonian
\begin{eqnarray}
\label{eq:SC_Bulk}
H_{\text{SC}}
=
\left[-\mu + m + t \sum_i \cos k_i\right]\xi_z
+
\Delta\,\xi_y \sigma_y .
\end{eqnarray}
Here, $\xi_i$ denote Pauli matrices in Nambu (particle-hole) space and $\sigma_i$ act on the spin degree of freedom. The parameter $\mu$ is the common chemical potential of the coupled TI-SC system.
We keep $m$ explicit so that the SC band position can be tuned independently without abandoning a common chemical-potential convention across the TI-SC stack.
The parameter $t$ sets the lattice-regularized normal state dispersion, and $\Delta$ is the uniform onsite $s$-wave pairing amplitude.
Near $\Gamma$, this model describes an approximately parabolic spin-degenerate band with a uniform $s$-wave gap $\Delta$, and is therefore topologically trivial before coupling to the TI [Fig.~\ref{fig2:SetUp}(c)].
Time-reversal symmetry acts as $T = -i \sigma_y K$, and inversion is represented by $P = \tau_z$ in the orbital subspace, where $\tau_i$ are Pauli matrices in orbital space.

We next bring the TI and SC into contact along the $z$ direction. In the resulting heterostructure, the bottom TI surface hybridizes locally with the top SC layer, while translation symmetry remains intact in the in-plane directions [Fig.~\ref{fig1:Schematics}(c)]. The corresponding heterointerfacial BdG Hamiltonian takes the block form
\begin{equation}
\label{eq:FullTISCHam}
H_{\text{HI}}(\mathbf{k})
=
\begin{pmatrix}
H_{\text{SC}}^{S}(\mathbf{k}) & H_{I} \\
H_{I}^{\dagger} & H_{\text{TI}}^{S}(\mathbf{k})
\end{pmatrix},
\end{equation}
where $H_{\text{TI}}^{S}$ denotes the low-energy Dirac BdG theory of the TI surface, while $H_{\text{SC}}^{S}$ is a minimal two-dimensional BdG model representing the SC surface sectors relevant for hybridization.
$H_I$ encodes the local interface tunneling.
For simplicity, we assume that only the two surfaces forming the buried heterointerface are directly coupled.

The low-energy surface Dirac cone on the TI side is described by the effective BdG Hamiltonian
\begin{eqnarray}
\label{eq:TI_surface}
H_{\text{TI}}^S (\mathbf{k}) = -\mu  \xi_z - \alpha k_x  \xi_z\sigma_y - \alpha k_y \sigma_x ,
\end{eqnarray}
where $\xi_i$ acts in Nambu space and $\sigma_j$ in spin space~[see Sec.~S1.1 for details]. 
This Hamiltonian captures a single Dirac cone with spin-momentum locking in the plane of the interface. 

We model the SC surface as a simple two-dimensional Hamiltonian and take the interface coupling to be a local hybridization $H_I = \lambda  \xi_z$, which mixes electron and hole sectors across the interface. 
Integrating out the SC degrees of freedom using a Schrieffer-Wolff transformation~\cite{schrieffer1966relation}, we obtain the low-energy effective Hamiltonian for the TI-SC heterointerface,
\begin{eqnarray}
\label{eq:TI_SW}
\bar{H}_{\text{TI}}^S (\mathbf{k}) &=& -\mu \xi_z - \alpha k_x \xi_z\sigma_y - \alpha k_y \sigma_x \nonumber \\
&& + \Delta_s  \xi_y \sigma_y + \Delta_p \left( k_x \xi_y + k_y \xi_x \sigma_z \right) ,
\end{eqnarray}
where $\Delta_s$ and $\Delta_p$ represent proximity-induced $s$-wave and effective $p$-wave pairings
on the TI-SC interface, respectively.
To leading nontrivial order in the interface hybridization,
\begin{eqnarray}
\Delta_s
&=&
\frac{\Delta\lambda^2}{(\mu+m+3t)^2-\mu^2},
\\
\Delta_p
&=&
\frac{2\Delta\lambda^2\mu\alpha}{\left[(\mu+m+3t)^2-\mu^2\right]^2},
\end{eqnarray}
so that the induced pairing structure and the associated interfacial gap follow directly from the microscopic TI-SC coupling rather than being introduced phenomenologically [see Sec.~S1.2].

The effect of the proximity coupling is illustrated in Figs.~\ref{fig2:SetUp}(d,e). 
In the absence of interface hybridization ($\lambda = 0$), the TI-SC heterostructure band structure simply consists of the ungapped Dirac surface state (blue) and the decoupled SC bands~[Fig.~\ref{fig2:SetUp}(d)]. 
Upon turning on a finite interface coupling, $\lambda = 3$ meV, the Dirac cone at the heterointerface acquires a superconducting gap (red) due to the induced pairing, which we denote by $\Delta_\text{ind}$~[Fig.~\ref{fig2:SetUp}(e)]. 
For representative parameters $m_0 = -120$~meV, $t_0 = 48$~meV, $\alpha = 40$~meV, $m = -45$~meV, $t = 30$~meV, $\Delta = 2$~meV, and $\mu =1.5$~meV, with TI thicknesses $l_T = 8$ and $l_S = 8$, the interface spectrum is fully gapped and adiabatically connected to the low-energy effective model in Eq.~\eqref{eq:TI_SW}.
For this choice of parameters, the induced gap extracted from Fig.~\ref{fig2:SetUp}(e) is $\Delta_\text{ind} \approx 0.355$~meV, significantly smaller than the parent pairing amplitude $\Delta$, as expected for a proximity-induced gap.
Note that although $\lambda$ seems large, the resulting induced gap indicates that it acts as a perturbative parameter~[see Eq.~(S13)].
The induced pairing $\Delta_\text{ind}$ contains an $s$-wave component $\Delta_s$ and a symmetry-allowed momentum-linear correction $\Delta_p$. The resulting buried heterointerface belongs to symmetry class DIII~\cite{schnyder2008classification,chiu2016classification} and lies in the topological regime $\Delta_s^2 \alpha^2 > \Delta_p^2 \mu^2$, 
as shown in Sec.~S1.2. 
In particular, the Fu-Kane limit, where $\Delta_p$ goes to zero, remains topological, while a finite $\Delta_p$ modifies the edge dispersion and the detailed structure of the Majorana wave functions.

We derive the effective edge theory of the heterointerface and obtain analytic expressions for the Majorana edge modes to clarify this TSC state.
For an interface that is finite along $x$ and $y$ with open boundaries, we distinguish edges by their outward normal. 
We label $i=1$ ($i=2$) as the edges normal to $y$ ($x$), so that the corresponding low-energy modes propagate along $x$ ($y$). In this notation, the edge Hamiltonian takes the form 
\begin{eqnarray}
H_i^\text{edge}(k_i) = (-1)^i k_i \left( \alpha  \xi_y + \Delta_p  \xi_x \right),
\end{eqnarray}
where $\xi_i$ is the effective Nambu space, and the dispersion is
\begin{eqnarray}
\varepsilon(k_i) = \pm k_i \sqrt{\alpha^2 + \Delta_p^2} .
\end{eqnarray}
In the projected low-energy subspace of a single helical Dirac channel, the edge theory reduces to an effective two-component Nambu spinor.
The resulting edge eigenstates satisfy the Majorana constraint $\psi=\mathcal{C}\psi$ with $\mathcal{C}=\xi_x K$ (up to a basis convention).
The corresponding edge-localized eigenstates can be written as
\begin{eqnarray}
\label{eq:Maj_WF}
\psi_1^\text{edge} =
\begin{pmatrix}
e^{i\theta} \\
e^{-i\theta}
\end{pmatrix} , \qquad
\psi_2^\text{edge} =
\begin{pmatrix}
e^{i\phi} \\
e^{-i\phi}
\end{pmatrix} ,
\end{eqnarray}
where the phases are defined by
\begin{eqnarray}
\theta = \tan^{-1} \frac{\Delta_p - \sqrt{\Delta_p^2 + \alpha^2}}{\Delta_p} , \qquad
\phi = \frac{\pi}{2} + \theta .
\end{eqnarray}
These boundary solutions are consistent with the time-reversal-invariant topological character of the proximitized heterointerface and with the helical Majorana edge modes~[see Sec.~S1.2 for a full derivation].

\begin{figure*}[t]
\includegraphics[width=0.99\textwidth]{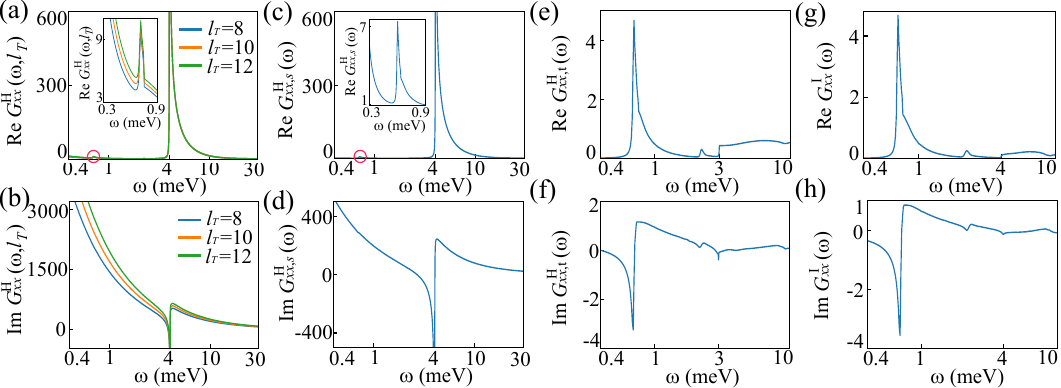}
\caption{\label{fig:opt_tisc}
\textbf{Optical response of the TI-SC heterostructure and isolation of the interface sheet conductance.}
\textbf{(a,e)} Real and imaginary parts of the total in-plane sheet conductance $G^{\text{H}}_{xx}(\omega,l_{T})/G_{0}$ of the TI-SC heterostructure for several TI thicknesses $l_{T}$. 
The spectra contain the TI bulk Drude response and Dirac surface contribution together with the SC superfluid and quasiparticle terms, and they display a pronounced feature near $\omega \simeq 2\Delta_{\text{ind}}$ associated with the proximitized interface states (highlighted in the insets, red circles). 
\textbf{(b,f)} Thickness-independent sheet component $G^{\text{H}}_{xx,s}(\omega)/G_{0}$ obtained by extrapolating $G^{\text{H}}_{xx}(\omega,l_{T})/G_{0}$ as $l_{T}$ goes to zero, thereby removing the TI bulk background that scales with thickness and retaining the combined sheet response of the top TI surface, the buried interface, and the SC slab. 
\textbf{(c,g)} Normalized surface sheet conductance $G^{\text{H}}_{xx,t}(\omega)/G_{0}$ after subtracting the calibrated SC sheet and bulk contributions, so that the remaining response arises from the top TI surface and the buried interface. 
\textbf{(d,h)} Interface-only sheet conductance $G^{\text{I}}_{xx}(\omega)/G_{0}$ obtained by further subtracting the known TI top-surface spectrum from the surface term in (c,g). 
This response consists of a single, sharply defined resonance at $\omega \simeq 2\Delta_{\text{ind}}$, identifying the induced coherence peak as an intrinsic property of the proximitized interface states rather than of TI or SC bulk channels. 
Model parameters are the same as in Fig.~\ref{fig2:SetUp}.}
\end{figure*}

\subsection*{Optical Response of the TI-SC Heterostructure}

We begin by calibrating the optical responses of the isolated TI and SC slabs, which provide the reference baselines for the buried-interface extraction in the full heterostructure.  
For the TI slab, the relevant low-energy scale is the chemical potential $\mu$ measured from the surface-Dirac point [Fig.~S2 and Sec.~S2.1]. 
The real part $\text{Re}\,G_{xx}^{\text{TI}}(\omega)$ contains a low-frequency Drude contribution together with a broader interband background arising from bulk and surface bands.
Superimposed on this background, the Dirac surface state produces a characteristic Pauli-blocked onset at $\omega=2|\mu|$: for $\omega<2|\mu|$, interband transitions across the surface Dirac cone are suppressed, whereas for $\omega>2|\mu|$ they become allowed and generate a step-like increase in the conductance.
The imaginary part $\text{Im}\,G_{xx}^{\text{TI}}(\omega)$ shows the corresponding inductive low-frequency response associated with the Drude term, together with a dispersive kink near $\omega\simeq 2|\mu|$ linked to the onset of Dirac interband absorption.
Both the Drude weight and the broad interband background increase with TI slab thickness $l_T$, indicating their dominant bulk origin [Figs.~S2(a,d)].
Treating $l_T$ as a control parameter, we fit the sheet conductance to
\begin{equation}
G_{xx}^{\text{TI}}(\omega,l_T)
=
G_{xx,s}^{\text{TI}}(\omega)
+
\sigma_{xx}^{\text{TI}}(\omega)\,l_T ,
\label{eq:ss_TI_fit}
\end{equation}
thereby separating a thickness-independent sheet term $G_{xx,s}^{\text{TI}}(\omega)$ [Figs.~S2(b,e)] from the effective bulk conductivity $\sigma_{xx}^{\text{TI}}(\omega)$.
As an independent validation, a layer-projected surface-only optical channel reproduces the same Pauli-blocked threshold and closely matches the extrapolated sheet response~[Figs.~S2(c,f)], confirming that the Dirac surface contribution behaves as a genuine two-dimensional conducting sheet superimposed on a bulk background.

We next analyze the optical response of an isolated superconducting slab~[Fig.~S3 of Sec.~S2.2].
In this case, the characteristic energy scale is set by the Bogoliubov-de Gennes gap $\Delta$ of the parent $s$-wave SC.
In the clean-limit BCS picture, the real part $\text{Re}\,G_{xx}^{\text{SC}}(\omega)$ exhibits an optical gap for $0<\omega\lesssim 2\Delta$, with vanishing regular absorption below $2\Delta$ and a pronounced coherence peak at $\omega\simeq 2\Delta$ where pair-breaking excitations first become allowed.
For $\omega>2\Delta$, the conductance evolves into a gapped quasiparticle continuum.
The imaginary part $\text{Im}\,G_{xx}^{\text{SC}}(\omega)$ shows the characteristic superconducting response, with a pronounced low-frequency $1/\omega$ divergence from the superfluid stiffness and a dispersive feature near $\omega \simeq 2\Delta$ associated with the coherence peak.
Both the superfluid weight and the quasiparticle continuum increase with the superconducting slab thickness $l_S$, confirming their bulk origin [Figs.~S3(a,d)].

As for the TI slab, both the superfluid weight and the quasiparticle continuum scale with the SC thickness $l_S$, allowing us to write [see more details in Sec.~S3]
\begin{equation}
G_{xx}^{\text{SC}}(\omega,l_S)
=
G_{xx,s}^{\text{SC}}(\omega)
+
\sigma_{xx}^{\text{SC}}(\omega)\,l_S ,
\label{eq:ss_SC_from_fit}
\end{equation}
which yields a thickness-independent sheet term $G_{xx,s}^{\text{SC}}(\omega)$ [Figs.~S3(b,e)] and a bulk conductivity $\sigma_{xx}^{\text{SC}}(\omega)$ from the linear slope [Figs.~S3(c,f)].

Having established these standalone responses,
we now turn to the full TI-SC heterostructure, in which the TI slab is stacked on top of the SC~[Fig.~\ref{fig:opt_tisc}]. 
The total in-plane sheet conductance $G_{xx}(\omega,l_{T})$ again displays a low-frequency Drude contribution together with a broad bulk-like background, both of which increase with the TI thickness $l_{T}$, as expected from the growing TI volume~[Figs.~\ref{fig:opt_tisc}(a,e)].
Superimposed on these contributions, a new low-frequency resonance emerges near the induced gap $2\Delta_{\text{ind}}$ of the proximitized interface states, as highlighted in the inset of Fig.~\ref{fig:opt_tisc}(a).

To isolate the buried-interface response, we again exploit the thickness scaling and fit.
For each frequency and fixed $l_S$, we treat $l_T$ as a control parameter and fit the heterostructure sheet conductance $G^{\text{H}}_{xx}(\omega, l_T)$ to the linear form
\begin{equation}
\label{eq:sigma_inter_fit_tisc_lin}
G^{\text{H}}_{xx}(\omega, l_T)
= G^{\text{H}}_{xx,s}(\omega)
+ \sigma^{\text{H}}_{xx}(\omega)\, l_T ,
\end{equation}
which separates the thickness-independent sheet contribution $G^{\text{H}}_{xx,s}(\omega)$ from the bulk background $\sigma^{\text{H}}_{xx}(\omega)$.  
Note that this linear approximation is sufficient in the frequency window of interest because residual nonlinear thickness corrections are negligible [see Methods and Table~S1]. 
Because the superconducting thickness $l_S$ is fixed, the extrapolated sheet term $G^{\text{H}}_{xx,s}(\omega)$ still contains the conventional $s$-wave superconducting response of the parent SC [see Sec.~S2.2 and Figs.~S3(b,e)].
As shown in Fig.~\ref{fig:opt_tisc}(b), the real part $\text{Re}\,G^{\text{H}}_{xx,s}(\omega)$ is strongly suppressed for $\omega < 2\Delta_{\text{ind}}$, and develops a clear, thickness-independent peak at $\omega \simeq 2\Delta_{\text{ind}}$.  
This feature is well separated from the broader parent $s$-wave gap structure appearing at $\omega = 2\Delta$.  
The imaginary part, $\text{Im}\,G^{\text{H}}_{xx,s}(\omega)$, combines the $1/\omega$ tail of the SC superfluid and residual Drude carriers with dispersive anomalies at these characteristic energies~[Fig.~\ref{fig:opt_tisc}(f)].

Using the calibrated optical responses of the standalone TI and SC slabs, we isolate the thickness-independent sheet contribution associated with the proximitized TI-SC interface. 
We first subtract $\tfrac{1}{2}G_{xx,s}^{\text{SC}}(\omega)$ and $\sigma^{\text{SC}}_{xx}(\omega)\,l_S$ using Eq.~\eqref{eq:ss_SC_from_fit} (with $l_S=8$ fixed), which is denoted by $G^{\text{H}}_{xx,t}(\omega)/G_{0}$, as shown in Figs.~\ref{fig:opt_tisc}(c,g)~[see also Figs.~S3(c,f)] to remove the SC outer surface contribution and SC bulk contribution.
We then remove the TI outer surface sheet response 
$\tfrac{1}{2}G^{\text{TI}}_{xx,s}(\omega)$.
As shown in Figs.~\ref{fig:opt_tisc}(d,h), the resulting residual interface-only sheet conductance $G^{\text{I}}_{xx}(\omega)$ is dominated by a sharply defined coherence peak at $\omega \simeq 2\Delta_{\text{ind}}$, which we identify as the optical fingerprint of the superconducting heterointerface.
A peak near $2\Delta_{\text{ind}}$ is already visible in the raw $G^{\text{H}}_{xx}(\omega,l_T)$ for each $l_T$ [Fig.~\ref{fig:opt_tisc}(a) inset]. Thus, the protocol above isolates its interface-localized optical response~[see the Methods section].

\begin{figure}[b]
\includegraphics[width=0.46\textwidth]{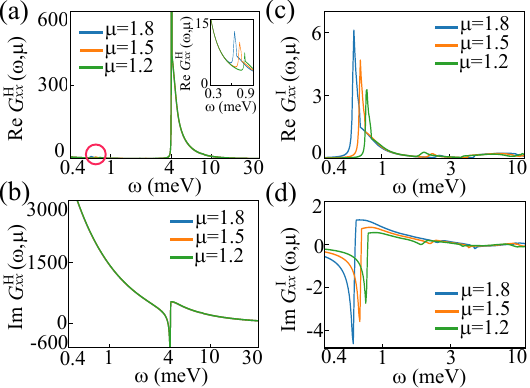}
\caption{\label{fig:opt_mu}
\textbf{Chemical-potential dependence of the interface optical response.}
\textbf{(a,b)} Real and imaginary parts of the total in-plane conductance $G^{\text{H}}_{xx}(\omega,\mu)/G_{0}$ for several chemical potentials $\mu$.  
The dominant feature at $\omega \simeq 2\Delta$ reflects the parent
$s$-wave condensate, while the inset in (a) resolves the additional
low-energy resonance associated with the induced interface gap (red circle).  
\textbf{(c,d)} Interface-only sheet conductance
$G^{\text{I}}_{xx}(\omega,\mu)/G_{0}$ obtained by subtracting the calibrated TI-surface and bulk-SC backgrounds from the total response.
The coherence peak near $2\Delta_{\text{ind}}$ shifts and reshapes with $\mu$, revealing the sensitivity of the induced pairing on the Dirac surface state to carrier density.  
All other parameters are as in Fig.~\ref{fig2:SetUp}.}
\end{figure}
This extraction procedure makes explicit that the induced-gap resonance is confined to the thickness-independent sheet channel localized at the buried interface. 
The pure TI slab fixes the outer-surface Dirac response, including its characteristic onset at $2|\mu|$, while the pure SC slab constrains the parent-gap feature near $2\Delta$ together with its accompanying bulk background.
Once these surface and bulk contributions are removed, the remaining interface-only sheet conductance is characterized by the single peak at $2\Delta_{\text{ind}}$. 
In contrast, the Drude component and broad bulk continua scale with thickness and are therefore eliminated in the extracted interfacial channel.
This interpretation is further supported by benchmarking the interface-only conductance across three complementary descriptions: the full slab calculation, a projected interface-band subspace, and the Schrieffer-Wolff low-energy effective theory~[Sec.~S1.3 and Fig.~S1].
All three approaches reproduce the same induced-gap coherence feature and comparable low-frequency lineshape, confirming that the extracted resonance is an intrinsic property of the proximitized interface sector rather than an artifact of a particular approximation.
\begin{figure}[t]
\includegraphics[width=0.46\textwidth]{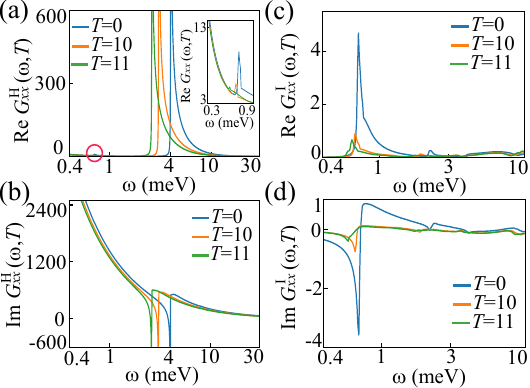}
\caption{\label{fig:opt_T}
\textbf{Temperature dependence of the interface optical response.}
\textbf{(a,b)} Real and imaginary parts of the total in-plane sheet conductance $G^{\text{H}}_{xx}(\omega,T)/G_{0}$ of the TI-SC heterostructure for several temperatures $T$. 
With increasing temperature, the structure associated with the parent $s$-wave gap near $\omega \simeq 2\Delta$ shifts to lower energies, while the interface-induced resonance near $\omega \simeq 2\Delta_{\text{ind}}$ is progressively suppressed.
The inset in (a) magnifies the induced-gap region marked by the red circle. 
\textbf{(c,d)} Interface-only sheet conductance $G^{\text{I}}_{xx}(\omega,T)/G_{0}$ obtained by subtracting the calibrated TI-surface and SC contributions. 
A single, sharp peak at $\omega \simeq 2\Delta_{\text{ind}}$ dominates the low-temperature spectra and is rapidly weakened and broadened on approaching $T_{c}$, underscoring that the interfacial coherence peak follows the superconducting order parameter while remaining at a parametrically smaller energy scale than the parent gap.
All other parameters are as in Fig.~\ref{fig2:SetUp}.}
\end{figure}
We now examine how the optical signature of the proximitized TI-SC interface evolves with chemical potential and temperature. 
Figures~\ref{fig:opt_mu} and \ref{fig:opt_T} summarize the behavior of both the total sheet conductance and the interface-isolated channel. 
Varying the chemical potential $\mu$ redistributes spectral weight in the total response through changes in the Dirac threshold and carrier density, while the extracted interfacial conductance shows that the coherence peak near $\omega \simeq 2\Delta_{\text{ind}}$ shifts and reshapes with $\mu$~[Fig.~\ref{fig:opt_mu}]. 
This identifies the induced interfacial scale as a tunable property of the proximitized surface band, whereas the parent superconducting feature near $\omega \simeq 2\Delta$ remains essentially unchanged over the same parameter range. 

By contrast, temperature primarily controls the overall superconducting energy scales. 
As shown in Figs.~\ref{fig:opt_T}(a,b), increasing $T$ shifts the parent $s$-wave structure near $\omega \simeq 2\Delta$ to lower energies and suppresses the low-frequency imaginary response associated with superfluid stiffness. 
In the extracted interface channel [Figs.~\ref{fig:opt_T}(c,d)], the resonance near $\omega \simeq 2\Delta_{\text{ind}}$ is progressively weakened and broadened, becoming indistinct as $T$ goes to $T_c$. 
Thus, while the induced interfacial scale is tunable with $\mu$, its temperature dependence remains locked to the parent superconducting order parameter and stays well separated from the bulk pair-breaking scale.

\begin{figure}[t]
\includegraphics[width=0.48\textwidth]{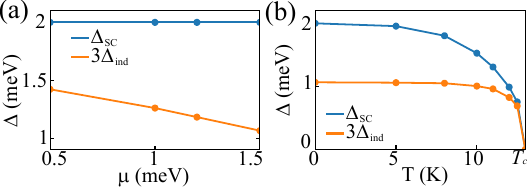}
\caption{\label{fig:gap_scales}
\textbf{Evolution of the parent and induced superconducting energy scales with chemical potential and temperature.}
\textbf{(a)}
Characteristic frequencies $\omega_{\text{SC}}(\mu)$ and $3\omega_{\text{ind}}(\mu)$ obtained from the peak positions of $\text{Re}\,G_{xx}^{\text{H}}(\omega,\mu)$, representing the parent superconducting gap and the induced TI-SC interface gap, respectively, as functions of chemical potential.
Here, the induced scale is multiplied by a factor of 3 only for visual clarity, so that its evolution can be compared directly with $\omega_{\text{SC}}$ on the same vertical axis; this rescaling has no physical significance.
\textbf{(b)}
Same analysis as a function of temperature at fixed $\mu$, showing $\omega_{\text{SC}}(T)$ and $3\omega_{\text{ind}}(T)$ extracted from $\text{Re}\,G_{xx}^{\text{H}}(\omega,T)$.
The two panels highlight that the induced interfacial scale is tunable and remains well separated from the parent superconducting scale while collapsing near $T_c$.
All other parameters are as in Fig.~\ref{fig2:SetUp}.}
\end{figure}

These trends are summarized in Fig.~\ref{fig:gap_scales}. 
The parent scale $\omega_{\text{SC}}$ is nearly independent of $\mu$, whereas the induced scale $\omega_{\text{ind}}$ varies substantially with carrier density. 
As a function of temperature, however, both scales decrease monotonically and collapse near $T_c$, confirming that the interfacial resonance is a proximity-induced superconducting scale rather than an independent low-energy feature.

\begin{figure*}[t]
\includegraphics[width=0.75\textwidth]{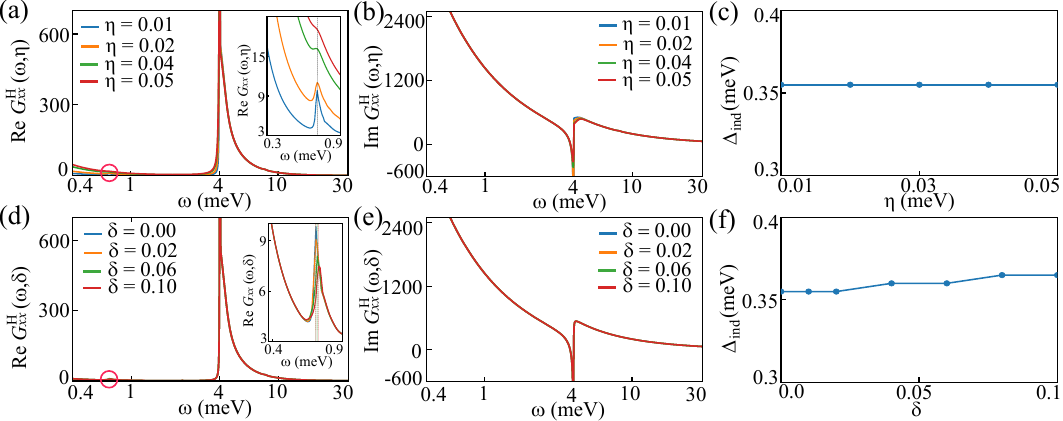}
\caption{\label{fig:disorder_eta_delta}
\textbf{Optical response of the TI-SC heterostructure under lifetime broadening and interface inhomogeneity.}
\textbf{(a,b)} Real and imaginary parts of the total in-plane sheet conductance $G^{\text H}_{xx}(\omega)$ for several phenomenological broadenings $\eta$.
Increasing $\eta$ broadens the spectral structures and reduces their amplitudes, while the low-energy interfacial feature near $\omega \simeq 2\Delta_{\rm ind}$ remains visible [highlighted in the inset of panel (a)].
The prominent structure near $\omega \simeq 2\Delta$ reflects the parent-SC pair-breaking/coherence scale.
\textbf{(c)} Induced gap $\Delta_{\rm ind}$ extracted from the interfacial band dispersion as a function of $\eta$.
The gap remains essentially unchanged, consistent with the fact that $\eta$ enters only as a linewidth parameter in the Kubo response.
\textbf{(d,e)} Real and imaginary parts of $G^{\text H}_{xx}(\omega)$ in the presence of interface inhomogeneity modeled by $\lambda_i=\lambda(1+\xi_i)$ with $\xi_i\in[-\delta,\delta]$.
Moderate inhomogeneity broadens the interfacial response and slightly renormalizes its amplitude, while the feature near $\omega \simeq 2\Delta_{\rm ind}$ remains clearly identifiable [see inset of panel (d)].
\textbf{(f)} Induced gap $\Delta_{\rm ind}$ as a function of the inhomogeneity strength $\delta$.
Only weak renormalization is observed over the range shown, indicating that moderate interface disorder does not obscure the induced-gap scale.
Model parameters are the same as in Fig.~\ref{fig2:SetUp} unless noted otherwise.
}
\end{figure*}

Finally, we briefly assess the robustness of the proposed extraction protocol against experimentally relevant disorder effects.
In realistic TI-SC heterostructures, finite quasiparticle lifetimes and imperfect buried interfaces can broaden optical features and introduce spatial variations in the local TI-SC transparency~\cite{Zimmermann1991,Blonder1982,PannetierCourtois2000}.
To capture these effects at a minimal level, we consider two complementary disorder proxies. The first is a phenomenological linewidth broadening $\eta$ introduced in the Kubo response [Eq.~\eqref{eq:Method_Cond}].
The second is interface inhomogeneity in the local TI-SC hybridization, modeled as $\lambda_i=\lambda(1+\xi_i)$, where $i$ labels the in-plane sites at the buried TI-SC interface, $\xi_i\in[-\delta,\delta]$, and $\delta$ controls the strength of the relative interface inhomogeneity.
As shown in Fig.~\ref{fig:disorder_eta_delta}(a,b,d,e), both effects broaden the interfacial lineshape and reduce its amplitude in the real and imaginary parts of the conductance, but over the range studied they do not eliminate the coherence feature near $\omega\simeq 2\Delta_{\rm ind}$ or obscure the induced-gap scale.
In particular, because $\eta$ enters only as a linewidth parameter in the optical kernel, the extracted $\Delta_{\rm ind}$ remains essentially unchanged [Fig.~\ref{fig:disorder_eta_delta}(c)], whereas moderate interface inhomogeneity produces only weak renormalization of the induced gap for $\delta \lesssim 0.1$ [Fig.~\ref{fig:disorder_eta_delta}(f)].
The thickness-extrapolation protocol therefore remains meaningful in the experimentally relevant regime of moderate lifetime broadening and moderate interface inhomogeneity.
Further details are provided in Sec.~S4.

Taken together, Figs.~\ref{fig:opt_tisc}-\ref{fig:disorder_eta_delta} establish a coherent picture of the proximity-induced superconducting state at the TI-SC interface.
The optical sheet conductance allows us to separate bulk and interfacial contributions, identify the thickness-independent Dirac-derived sheet channel, and track the opening of an induced gap that is distinct from the parent superconducting gap.
The evolution of the corresponding energy scales with chemical potential and temperature shows that the induced gap is tunable by carrier density and remains tied to the parent order parameter through its temperature dependence, vanishing at the same critical temperature.
These results demonstrate that terahertz spectroscopy can resolve the hierarchy of superconducting energy scales in TI-SC heterostructures and provide a noninvasive, bulk-compatible probe of buried proximitized interfaces that is complementary to conventional surface-sensitive techniques.

\begin{figure*}[t]
    \includegraphics[width=0.98\textwidth]{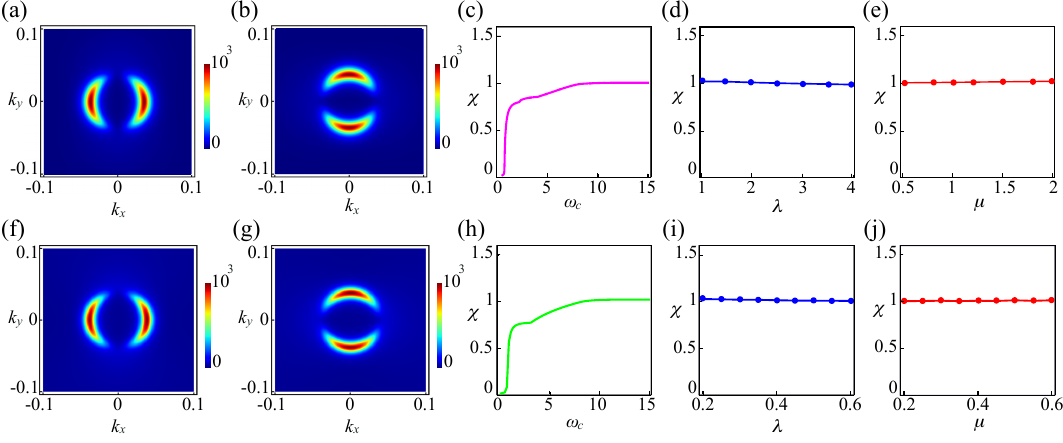}
    \caption{\label{fig:quantumweight}
    \textbf{Quantum metric and optical sum rule at the TI-SC interface.} 
    \textbf{(a-e)} Full TI-SC heterostructure results from the projected interface-bands subspace. 
    (a,b) Quantum-metric components $g_{xx}(\mathbf{k})$ and $g_{yy}(\mathbf{k})$, strongly enhanced on the minimal-gap ring of the proximitized interface. 
    (c) Optics-geometry ratio $\chi(\omega_c)$ [Eq.~\eqref{eq:chiratio}] obtained from the interface-only sheet conductance $G_{xx}(\omega)$ and the quantum-metric integral over the projected interface-bands subspace. 
    $\chi(\omega_c)$ rises sharply at the induced-gap coherence peak $\omega=2\Delta_{\text{ind}}\simeq0.71~\text{meV}$ and saturates to $\chi\simeq1$ at higher cutoff.  
    (d,e) $\chi$ versus proximity coupling $\lambda$ and chemical potential $\mu$, evaluated over $\mathcal{R}:(k_x,k_y)\in[-0.1,0.1]\times[-0.1,0.1]$, showing $\chi\simeq1$ across moderate parameter variations. In (d) $\lambda$ is varied keeping $\mu$ fixed at $1.5\ \text{meV}$ and in (e) $\mu$ is varied with proximity coupling $\lambda = 3$.
    \textbf{(f-j)} Schrieffer-Wolff effective interfacial BdG theory. 
    (f,g) Quantum-metric maps exhibiting the same minimal-gap-ring enhancement as in (a,b). 
    (h) $\chi(\omega_c)$ computed fully within the effective theory, reproducing the onset at $2\Delta_{\text{ind}}$ and saturation to $\chi\simeq1$ in (c).  
    (i,j) $\chi$ versus $\lambda$ and $\mu$ over the same region $\mathcal{R}$, remaining close to unity. In (i) $\lambda$ is varied keeping $\mu$ fixed at $0.6\ \text{meV}$ and in (j) $\mu$ is varied with a fixed $\lambda = 0.6$. All other model parameters are the same as in Fig.~\ref{fig2:SetUp}.
    }
\end{figure*}

\subsection*{Quantum Geometry from the Optical Response}

Having established that the interface sheet conductance exhibits a coherence peak at $\omega = 2\Delta_{\text{ind}}$, we now examine how the associated low-frequency optical spectral weight is related to the quantum geometry of the proximitized interface state.
We show that, once the buried interfacial contribution has been isolated, its generalized optical weight admits a compact quantum-geometric interpretation.

In a generic gapped multiband system, the linear-response optical conductivity/conductance is controlled not only by band dispersion but also by the underlying quantum geometry of the Bloch (or BdG) bands, encoded in the quantum metric~\cite{provost1980riemannian,10.1093/nsr/nwae334}. 
For the effective two-dimensional interface Hamiltonian in Eq.~\eqref{eq:TI_SW}, we characterize this geometry by the quantum weight $K_{\mu \nu} = 2\pi \int_{BZ} \frac{d^{2}\mathbf{k}}{4 \pi^{2}}g_{\mu\nu}(\mathbf{k})$, defined as the Brillouin-zone integral of the quantum metric $g_{\mu\nu}(\mathbf{k})$~\cite{PhysRevResearch.7.023158}. 
This coincides with the quadratic coefficient in the small-wave-vector expansion of the static structure factor and thus provides a geometric measure of the spatial fluctuations of the superconducting ground state. 
The quantum metric is defined as~\cite{provost1980riemannian}
\begin{equation}
    \label{eq:qmetric}
    g_{\mu\nu}(\mathbf{k})
    =
    \frac{1}{2}\text{Tr}\!\left[\partial_{\mu}P(\mathbf{k})\,\partial_{\nu}P(\mathbf{k})\right],
\end{equation}
where $P(\mathbf{k}) = \sum_{E_{n}(\mathbf{k})<0}|u_{n}(\mathbf{k})\rangle \langle u_{n}(\mathbf{k})|$  projects onto the occupied BdG quasiparticle subspace and $|u_{n}(\mathbf{k})\rangle$ is the $n$th normalized eigenvector of the BdG Hamiltonian.

The connection to the optical response follows from the fluctuation-dissipation relation between equal-time density correlations and the frequency-integrated dynamical structure factor, which constrains the longitudinal optical conductivity/conductance in terms of the quantum metric~\cite{PhysRevResearch.7.023158,PhysRevX.14.011052,PhysRev.83.34,souza2000polarization}.  
For the interface-only sheet conductance $G_{\mu\mu}(\omega)$, this gives the longitudinal sum rule~\cite{PhysRevResearch.7.023158},  
\begin{equation}
    \label{eq:conductivity_q_weight}
    \int_{0^{+}}^{\infty} d\omega \,\frac{\text{Re}\,G_{\mu \mu}(\omega)}{\omega}
    =
    \frac{\pi e^{2}}{h} \,K_{\mu \mu} \,.
\end{equation}
Thus, the negative-first moment of the optical conductance is a purely geometric quantity, fixed by the integrated quantum metric of the occupied interface bands.  

We evaluate this optics-geometry correspondence in two complementary descriptions of the TI-SC heterointerface.
We first extract the interfacial quantum metric and sheet conductance from the full TI-SC slab [Eq.~\eqref{eq:FullTISCHam}] by constructing a projected low-energy interface-bands subspace.
On the other hand, we also compute the same quantities within the Schrieffer-Wolff low-energy effective interfacial BdG Hamiltonian [Eq.~\eqref{eq:TI_SW}].
Their agreement provides a stringent validation that the optical resonance directly reflects the interfacial quantum metric. 

Figs.~\ref{fig:quantumweight}(a,b) report the momentum-resolved quantum-metric components $g_{xx}(\mathbf{k})$ and $g_{yy}(\mathbf{k})$ evaluated within this  projected interface-band subspace using full TI-SC slab. 
Both exhibit strong enhancement on a closed ring in momentum space where the direct BdG quasiparticle gap is minimal.
This ring-like structure reflects the generic inverse-squared gap dependence of the quantum metric and identifies the interfacial momentum region that dominates the low-frequency optical response.
 
Motivated by this structure, we define a low-energy momentum window $\mathcal{R}$ enclosing the minimal-gap ring and compare the corresponding partial quantum weight with the low-frequency generalized optical weight extracted from the interface-only sheet conductance $G_{xx}(\omega)$.
We introduce the dimensionless optics-geometry ratio $\chi(\omega_c)$, defined as the partial optical sum rule normalized by the quantum-metric integral over $\mathcal{R}$:
 \begin{equation}
\label{eq:chiratio}
\chi(\omega_c)=
\left[\int_{0^+}^{\omega_c} d\omega\,\frac{\text{Re}\,G_{xx}(\omega)}{\omega}\right]
\Bigg/
\left[\frac{e^{2}}{2h}\int_{\mathcal{R}} d^{2}\mathbf{k}\,g_{xx}(\mathbf{k})\right].
\end{equation}
By construction, $\chi(\omega_c)\simeq 1$ indicates that the low-frequency optical integral has nearly saturated the quantum-metric weight contained in $\mathcal{R}$.
As shown in Fig.~\ref{fig:quantumweight}(c), $\chi(\omega_c)$ rises sharply as the cutoff crosses the induced-gap coherence peak at $\omega = 2\Delta_{\text{ind}} \simeq 0.71~\text{meV}$ and then approaches $\chi \simeq 1$ at higher cutoff.
This shows that the dominant fraction of the interfacial quantum weight is already captured by the low-frequency optical spectral weight associated with the proximity-induced gap opening.

This correspondence is reproduced independently within the Schrieffer-Wolff effective theory and is therefore not an artifact of the projection procedure.
As shown in Figs.~\ref{fig:quantumweight}(f,g), the Schrieffer-Wolff effective interfacial BdG Hamiltonian yields the same minimal-gap-ring enhancement of the quantum metric found in the full TI-SC heterostructure [Figs.~\ref{fig:quantumweight}(a,b)].
Consistently, the Schrieffer-Wolff evaluation of $\chi(\omega_c)$ [Fig.~\ref{fig:quantumweight}(h)] reproduces the sharp onset near $2\Delta_{\text{ind}}$ and the subsequent saturation to $\chi \simeq 1$ seen in Fig.~\ref{fig:quantumweight}(c).
These results show that the low-frequency generalized optical weight of the buried interface admits a controlled quantum-geometric interpretation.

This conclusion remains robust under moderate variations of the proximity coupling $\lambda$ and chemical potential $\mu$.
In the full TI-SC slab, $\chi$ stays close to unity over the parameter ranges shown in Figs.~\ref{fig:quantumweight}(d,e),
and the same behavior is recovered within the Schrieffer-Wolff theory in Figs.~\ref{fig:quantumweight}(i,j), 
despite the expected renormalization of the effective low-energy parameters.
The persistence of $\chi \simeq 1$ in both descriptions establishes that the optical-sum-rule--quantum-metric correspondence is a generic low-energy property of the proximitized interface sector rather than a fine-tuned feature of a particular approximation or parameter set.

Consequently, integrating the interfacial generalized optical weight up to the induced-gap coherence peak provides a controlled lower bound that rapidly approaches the quantum-metric weight contained in $\mathcal{R}$.

\section*{Discussion}

\subsection*{Model Scope and Experimental Relevance}

The layered TI-SC slab model is intended as a minimal microscopic description of proximity-induced superconductivity at a buried topological interface, while remaining anchored to experimentally relevant scales.
On the TI side, we employ a lattice-regularized four-band model that reproduces the band inversion and single helical Dirac surface state of the Bi$_2$Se$_3$ family near $\Gamma$~\cite{zhang2009topological,liu2010model}.
It should therefore be viewed as a minimal Bi$_2$Se$_3$-type model rather than a material-specific full-band fit.
For numerical clarity, we compress the absolute TI bulk energy scales, while preserving the band inversion, the surface Dirac structure, and the hierarchy of scales relevant to buried-interface proximity effects.

The superconducting sector represents a metallic band with several-hundred-meV bandwidth and a meV-scale pairing amplitude, consistent with widely used TI-SC proximity platforms.
These scales are broadly compatible with Nb, NbN, NbSe$_2$, and Pb thin films, where meV-scale parent gaps and transition temperatures of order $10$ K are routinely realized~\cite{wang2012coexistence,zareapour2012proximity,trang2020conversion,xu2015experimental}.
The interface hopping is chosen in an intermediate-proximity regime such that $\Delta_{\text{ind}} \ll \Delta$, while the buried Dirac sector remains well defined and its induced-gap feature remains spectroscopically resolvable.

Once the TI thickness exceeds the top-bottom hybridization regime, typically beyond about $4$-$5$ quintuple layers, further increases in thickness do not qualitatively modify the buried-interface response.
In this regime, the induced gap and its optical coherence features are controlled primarily by the local TI-SC hybridization at the buried interface,
whereas additional TI thickness primarily adds noninterfacial background from the TI bulk and the exposed top surface.
These extra contributions are removed by the thickness-extrapolation protocol.

For the representative parameters used here, we obtain $\Delta_{\text{ind}} \approx 0.355$~meV and hence an optical feature near $2\Delta_{\text{ind}} \approx 0.71$~meV, directly in the experimentally accessible sub-THz/THz window.
This scale is consistent with induced gaps reported in several TI-based proximity platforms~\cite{wang2012coexistence,zareapour2012proximity,trang2020conversion,xu2015experimental}.
By choosing the chemical potential close to the Dirac point, we emulate gated or compensated Bi$_2$Se$_3$- or Bi$_2$Te$_3$-based samples with suppressed bulk carriers, which is also the regime in which the buried proximitized interface is expected to be most clearly isolated.
The predicted resonance should therefore be identifiable not only through its position near $2\Delta_{\text{ind}}$, but also through its thickness independence and its characteristic evolution with temperature in terahertz time-domain spectroscopy or infrared ellipsometry~\cite{valdes2012terahertz,di2012optical,tang2013terahertz}.

\subsection*{Limitations and Outlook}

Our analysis is formulated within a minimal, clean, and translationally invariant slab description, which defines the regime of validity of the proposed protocol.
In particular, we neglect strong disorder, interface roughness, and strain inhomogeneity, and model the superconducting layer within a single-band mean-field BCS framework. We also restrict attention to linear-response optics. 

Experimentally, implementing the thickness-extrapolation protocol requires controlled film thickness, chemical potential, and interface quality across a sample series, together with phase-sensitive measurements of the complex sheet conductance.
In practice, this relies on standard thin-film analysis of terahertz transmission or infrared ellipsometry data.
Although systematic uncertainties from multilayer optics, substrate response, and thickness calibration must be controlled carefully,
the required ingredients are within the reach of current epitaxial growth,
van der Waals assembly, and terahertz spectroscopy techniques.

As a complementary direction, Sec.~S5 analyzes the optical response of the helical Majorana boundary modes supported by the same proximitized interface.
The helical Majorana modes exhibit a low-frequency vertical absorption absent for ordinary helical fermionic edges, and under a weak Zeeman field their threshold behavior differs qualitatively from that of gapped quantum spin Hall edges~\cite{majoranaopticalcond}.
Since the relevant edge gap lies in the microwave regime, this points to microwave spectroscopy as a possible boundary-sensitive probe complementary to the buried-interface optical protocol.

More broadly, it will be interesting to extend the present approach to multiband or more strongly correlated topological superconducting platforms, including higher-order topological insulators, moir\'e flat-band systems, and quantum anomalous Hall-based heterostructures.
In such settings, the combination of interface-sensitive optics, thickness extrapolation, and quantum-geometric sum rules may provide a useful framework for characterizing both superconducting order and band topology in future topological quantum devices.

\section*{Conclusion}
We have shown that the buried TI-SC interface can be diagnosed optically through its thickness-independent sheet conductance.
By combining the Kubo formalism with thickness scaling to separate bulk and sheet channels,
we isolate an interface-only response whose dominant low-frequency resonance is set by the induced superconducting gap of the proximitized interface rather than by bulk TI or bulk SC contributions.
We further show that a quantum-metric sum rule governs this interfacial optical response,
such that the negative first moment of the conductance provides a controlled measure of the proximity-induced quantum weight of the gapped Dirac interface state.
The associated resonance evolves systematically with chemical potential and temperature,
and remains identifiable under moderate lifetime broadening and interface-coupling inhomogeneity over the range studied.
These results establish an optical route for identifying emergent topological superconductivity in TI-SC heterostructures, complementary to boundary-sensitive transport probes. 
The most direct experimental test is terahertz or infrared spectroscopy on TI-SC thin-film series with systematically varied TI thickness,
which should reveal a thickness-independent resonance at the induced interfacial gap scale and thereby distinguish a buried-interface superconducting feature from bulk optical backgrounds~\cite{lee2023gapless}. 
More broadly, magnetic-field tuning may provide a complementary handle on the associated Majorana sector by modifying the low-energy boundary spectrum and its optical response~\cite{sarma2015majorana,sanno2022engineering}.

\section*{Methods}

\subsection*{Model System Analysis}
To describe the low-energy electronic structure of the TI surface, we start from the linearized Dirac Hamiltonian expanded about the bulk gap-closing point at $\Gamma$~\cite{schindler2020dirac},
which captures the band inversion and spin-momentum locking of a three-dimensional TI.
This linearized Hamiltonian of the TI is split into the Dirac-like section $H^D_\text{TI}$ and the remainder $H^R_\text{TI}$, which are 
\begin{eqnarray}
H_{\text{TI}}^D &=& M\tau_z - i \alpha \tau_x \sigma_z \partial_z,\label{eq:TI_D}\\
H_{\text{TI}}^R &=& -\mu \tau_0 + \alpha \tau_x(k_x \sigma_x + k_y \sigma_y),\label{eq:TI_P}
\end{eqnarray}
where $\tau_{i}$ and $\sigma_{i}$ are Pauli matrices acting in orbital and spin space, respectively, and $M = m_0 + 3 t_0$ sets the bulk Dirac mass at the $\Gamma$ point.
The Fermi velocity of the Dirac cone is  given as $\alpha$.
The chemical potential $\mu$ is measured relative to the bulk band inversion and is chosen such that the Fermi level lies in the vicinity of the surface Dirac cone in the heterostructure.

The surface modes are obtained by solving the Dirac-like part $H_{\text{TI}}^D$ with a spatially varying mass $M(z)$ that changes sign across the TI-vacuum boundary.  
This reduces to a Jackiw-Rebbi problem, and yields exponentially localized eigenstates at the TI surface~\cite{jackiw1976solitons,jackiw1981solitons}.  
Projecting the remainder Hamiltonian $H_{\text{TI}}^R$ onto the subspace spanned by these Jackiw-Rebbi solutions leads to an effective two-dimensional surface Dirac Hamiltonian with Fermi velocity $\alpha$ and chemical potential $\mu$, as summarized in Sec.~S1.1.  

The superconducting slab is described by a minimal $s$-wave BdG model that is periodic in-plane and finite along the surface normal,
and we focus on the interfacial degrees of freedom relevant for proximity coupling to the TI surface sector.
Unlike the TI, the parent SC does not host a topological surface Dirac mode; hence, its impact on the low-energy surface response arises mainly through hybridization with the TI surface bands at the heterointerface.
To model this hybridization, we introduce a local tunneling Hamiltonian between TI and SC electrons,
\begin{eqnarray}
\label{eq:Hint_TI-SC}    H_{\text{I}}=\sum_{j,s}\lambda\left(a^\dagger_{j,s}c_{j,s}+c^\dagger_{j,s}a_{j,s}\right),
\end{eqnarray}
where $a_{j,s}$ and $c_{j,s}$ are the annihilation operators for the electrons of the SC and TI for spin $s$ at site $j$, respectively.
By applying the Schrieffer–Wolff transformation with the TI surface sector as the low-energy subspace and the coupled SC states as virtual high-energy excitations,
we obtain the dressed TI surface Hamiltonian in Eq.~\eqref{eq:TI_SW}.
It incorporates an induced pairing term and a renormalized Dirac velocity, serving as a minimal effective model for the proximitized surface [see more details in Sec.~S1.2].

Using the Jackiw-Rebbi solutions, we derive the edge Hamiltonians of the heterointerface Hamiltonian~\cite{schindler2020dirac},
\begin{eqnarray}
H_x^\text{edge} (k_x) &=&  -k_x\left(\alpha\xi_y+\Delta_p\xi_x\right)\label{eq:x_edge},\\
H_y^\text{edge} (k_y) &=&  k_y\left(\alpha\xi_y+\Delta_p\xi_x\right)\label{eq:y_edge},
\end{eqnarray}
where $H_i^\text{edge}(k_i)$ describes the edge along direction $i = x,y$ with conserved momentum $k_i$, and $\xi_{x,y}$ are Pauli matrices acting in the effective edge (Nambu or Majorana) pseudospin space.  
The resulting zero-energy states of the edge Hamiltonians of Eqs.~\eqref{eq:x_edge} and \eqref{eq:y_edge} are the Majorana edge states given in Eq.~\eqref{eq:Maj_WF}~[see more details in Sec.~S1.2].

\subsection*{Optical Response of Slab Geometries}
The in-plane optical response of the TI slabs, SC slabs, and TI-SC heterostructures is computed within a unified linear-response framework using a slab Hamiltonian $H_{\text{slab}}(\mathbf{k})$ that is periodic in the $(x,y)$ directions and open along $z$.
For each in-plane momentum $\mathbf{k}=(k_x,k_y)$, we diagonalize the normal or BdG slab Hamiltonian,
\begin{equation}
H_{\text{slab}}(\mathbf{k})\ket{\psi_{n\mathbf{k}}}=E_{n\mathbf{k}}\ket{\psi_{n\mathbf{k}}},
\end{equation}
with eigenstates normalized over the two-dimensional Brillouin zone.
This two-dimensional normalization implies that the current operator extracted from $H_{\text{slab}}$ naturally describes a current per unit length and thus yields a sheet conductance rather than a bulk conductivity.

The in-plane velocity operator is defined by the $k$-derivative of the slab Hamiltonian,
\begin{equation}
v_x(\mathbf{k}) = \frac{1}{\hbar}\partial_{k_x} H_{\text{slab}}(\mathbf{k}),
\end{equation}
and the corresponding current operator is
\begin{equation}
\hat{J}_x^{2\text{D}}(\mathbf{k}) = -ev_x(\mathbf{k}).
\end{equation}
Because $H_{\text{slab}}(\mathbf{k})$ is normalized per unit area, $\hat{J}_x^{2\text{D}}$ transports charge per unit length, in contrast to the three-dimensional current density (A/m$^2$) that appears in bulk Kubo formulae.
The linear-response kernel constructed from $\hat{J}_x^{2\text{D}}$ therefore has units $[G_{xx}] = \frac{\text{A/m}}{\text{V/m}} = \text{S},$ and we interpret $G_{xx}(\omega,l)$ as the in-plane sheet conductance of a slab of thickness $l$.

In the eigenbasis $\{\ket{\psi_{n\mathbf{k}}}\}$, the interband part of the longitudinal sheet conductance, $G_{xx}^{\text{Int}}(\omega,l)$, takes the standard form
\begin{equation}
\label{eq:Method_Cond}
G_{xx}^{\text{Int}}(\omega,l)=\frac{i}{N_k}\sum_{\mathbf k}\sum_{m\neq n}\frac{f_{n\mathbf k}-f_{m\mathbf k}}{\Omega_{mn}(\mathbf k)}\frac{\big|\langle \psi_{n\mathbf k} | \hat{J}_x^{2\text{D}} | \psi_{m\mathbf k}\rangle\big|^2}{\omega+i\eta-\Omega_{mn}(\mathbf k)},
\end{equation}
where $\Omega_{mn}(\mathbf{k})=E_{m\mathbf{k}}-E_{n\mathbf{k}}$, $f_{n\mathbf{k}}$ is the Fermi function, $N_k$ is the number of $\mathbf{k}$-points, and $\eta$ is a small phenomenological broadening.
For uniform $\mathbf{k}$ meshes, $(1/N_k)\sum_{\mathbf{k}}$ approximates the Brillouin-zone average.
No explicit factor of $l$ appears in this expression, since the current operator is already defined per unit length.
Any residual thickness dependence of $G_{xx}(\omega,l)$ therefore reflects a redistribution of spectral weight between surface-like and bulk-like layers rather than a change in the dimensionality of the response.
We evaluate the retarded response by the prescription $\omega\to\omega+i\eta$, and $\text{Re}\,G_{xx}(\omega)$ reported in the main text refers to the regular (finite-frequency) absorption, with the $\omega=0$ singular contribution encoded separately in the Drude term.

The full complex sheet conductance is written as
\begin{equation}
G_{xx}(\omega,l)=G^{\text{D}}_{xx}(\omega,l)+G^{\text{Int}}_{xx}(\omega,l),
\end{equation}
where $G^{\text{D}}_{xx}(\omega,l)=i D_{xx}(l)/(\omega+i\eta)$ is the Drude contribution characterized by a slab-thickness-dependent Drude weight $D_{xx}(l)$, and $G^{\text{Int}}_{xx}$ is the interband contribution.
In the superconducting (BdG) case, we avoid double counting of particle-hole degrees of freedom by including the standard overall factor of $1/2$ in the current-current response kernel.
Throughout we report $G_{xx}(\omega,l)$ in units of the conductance quantum $G_0=e^2/h$.

All calculations are converged with respect to the number of $\mathbf{k}$-points and the frequency sampling.
We use uniform meshes in the two-dimensional Brillouin zone with up to $\mathcal{O}(10^4)$ $\mathbf{k}$-points and choose $\eta$ and $T$ to match the values used in the main-text figures, thereby mimicking realistic disorder and finite-temperature broadening.

To separate bulk, surface, and interface contributions to the optical response, we exploit the real-space structure of the slab eigenstates.
For each eigenstate $\ket{\psi_{n\mathbf{k}}}$, we compute the probability weight on every atomic layer along $z$ and group these layers into predefined regions, top TI, bottom SC surfaces, TI bulk, interface layers, and SC bulk.
The corresponding region projectors are then used to reweight the Kubo kernel and generate channel-resolved sheet conductance, such as surface-surface, surface-bulk, and interface-interface responses, which is consistent with the layer-resolved analysis below.

\subsection*{Layer-Resolved Analysis}
To separate surface and interface responses from bulk contributions, we analyze the thickness scaling of the sheet conductance and obtain the thickness-independent term by extrapolation.
For each $(\mu,T,\eta)$ and frequency $\omega$, we compute the longitudinal sheet conductance $G_{xx}(\omega,l)$ for a series of slab thicknesses $l$ and fit it to a low-order polynomial,  
\begin{equation}
\label{eq:sigma_inter_fit_cp}
G_{xx}(\omega,l)=G_{xx,s}(\omega)+\sigma_{xx}(\omega)\,l+c(\omega)\,l^2,
\end{equation}
where $G_{xx,s}(\omega)$ is the thickness-independent sheet contribution dominated by surfaces and interfaces, $\sigma_{xx}(\omega)$ is an effective bulk conductivity setting the linear background, and $c(\omega)$ captures weak nonlinearities in $l$ due to finite-size and boundary effects.  
At each frequency $\omega$, we perform an independent fit over the thickness series used in the figure calculations. 
In practice, $c(\omega)$ is found to be negligible over the frequency range of interest, and a purely linear form $G_{xx}(\omega,l)\simeq G_{xx,s}(\omega)+\sigma_{xx}(\omega)l$ is sufficient in this work [see the representative fitting parameters in Table~S1].

For pure TI slabs, we apply Eq.~\eqref{eq:sigma_inter_fit_cp} to the TI sheet conductance $G^{\text{TI}}_{xx}(\omega,l_T)$,  
\begin{equation}
\label{eq:sigma_inter_fit_ti}
G^{\text{TI}}_{xx}(\omega,l_T)=G_{xx,s}^{\text{TI}}(\omega)+\sigma_{xx}^{\text{TI}}(\omega)\,l_T+c^{\text{TI}}(\omega)\,l_T^2,
\end{equation}
where $l_T$ is the TI thickness, $G_{xx,s}^{\text{TI}}(\omega)$ denotes the combined sheet conductance of the two TI surfaces, and $\sigma_{xx}^{\text{TI}}(\omega)$ is the effective TI bulk conductivity.  
Since there are two surfaces (top and bottom) for a pure TI slab, the single TI surface sheet conductance is  identified as  
\begin{equation}
\label{eq:ss_TI_single}
G_{xx,\text{single}}^{\text{TI}}(\omega)=\tfrac{1}{2}\,G_{xx,s}^{\text{TI}}(\omega).
\end{equation}
The sheet conductance extracted from the thickness series agrees quantitatively with the surface-only channel obtained from the layer projectors, providing an internal consistency check of the extrapolation procedure [see Figs.~S2(b,c,e,f) in Sec.~S2.1].

For pure SC slabs, we perform an analogous analysis by varying the SC thickness $l_S$ and fitting  
\begin{equation}
\label{eq:sigma_inter_fit_sc}
G^{\text{SC}}_{xx}(\omega,l_S)=G_{xx,s}^{\text{SC}}(\omega)+\sigma_{xx}^{\text{SC}}(\omega)\,l_S+c^{\text{SC}}(\omega)\,l_S^2,
\end{equation}
where $G_{xx,s}^{\text{SC}}(\omega)$ captures the thickness-independent sheet response of the two SC surfaces and $\sigma_{xx}^{\text{SC}}(\omega)$ is the SC bulk conductivity [see Fig.~S3 in Sec.~S2.2].
The single SC surface sheet conductance is then identified as  
\begin{equation}
\label{eq:ss_SC}
G_{xx,\text{single}}^{\text{SC}}(\omega)=\tfrac{1}{2}\,G_{xx,s}^{\text{SC}}(\omega).
\end{equation}
In practice, both $c^{\text{TI}}(\omega)$ and $c^{\text{SC}}(\omega)$ are negligible, ensuring that the purely linear form is sufficient~[see Table~S1].  

For TI-SC heterostructures, we fix the SC thickness $l_S$, vary the TI thickness $l_T$, and fit  
\begin{equation}
\label{eq:sigma_inter_fit_tisc}
G^{\text{H}}_{xx}(\omega,l_T)=G_{xx,s}^{\text{H}}(\omega)+\sigma_{xx}^{\text{H}}(\omega)\,l_T+c^{\text{H}}(\omega)\,l_T^2,
\end{equation}
to obtain the total thickness-independent sheet term $G_{xx,s}^{\text{H}}(\omega)$ and the effective bulk background $\sigma_{xx}^{\text{H}}(\omega)$ at each frequency.  
At fixed $l_S$, the $l_T$-independent intercept $G_{xx,s}^{\text{H}}(\omega)$ contains three contributions, the outer TI surface, the buried TI-SC interface, and the outer SC surface term, together with the fixed SC bulk background $\sigma_{xx}^{\text{SC}}(\omega)\,l_S$ associated with the finite superconducting thickness.
The outer TI surface is calibrated using the pure TI series via Eq.~\eqref{eq:ss_TI_single}, while the SC surface and bulk contributions are obtained from the pure SC series via Eqs.~\eqref{eq:sigma_inter_fit_sc}.  
Subtracting these baselines isolates the interface's spectrum,
\begin{eqnarray}
\label{eq:ss_interface_only_from_fit}
\nonumber
G_{xx}^{\text{I}}(\omega) &\simeq& G_{xx,s}^{\text{H}}(\omega)-G_{xx,\text{single}}^{\text{TI}}(\omega) \\
&&-G_{xx,\text{single}}^{\text{SC}}(\omega)
-\sigma_{xx}^{\text{SC}}(\omega)\,l_S,
\end{eqnarray}
which is the quantity plotted as the interface sheet conductance in Figs.~\ref{fig:opt_tisc}-\ref{fig:opt_T}. 
The approximation in Eq.~\eqref{eq:ss_interface_only_from_fit} reflects that we fix $l_S$ in the heterostructure series and treat the residual SC response as the calibrated single-surface term plus the bulk background $\sigma^{\text{SC}}_{xx}(\omega)l_S$. 

The same logic applies to the low-frequency Drude response.  
By fitting the Drude weight, or equivalently the zero-frequency limit of $\omega\,\text{Im}\,G_{xx}(\omega,l)$, as a function of $l$, a bulk contribution that scales with thickness $l$, can be separated from a thickness-independent surface or interface Drude term.

Note that the sheet conductance extracted from the thickness series agrees quantitatively with the surface-only channel obtained from the layer projectors, providing an internal consistency check of the extrapolation procedure.

\section*{Acknowledgments}
M. Kang, Y. Prasad, N.D. Babu, R. Ghadimi, and S. Cheon were supported by the National Research Foundation of Korea (NRF), funded by the Ministry of Science and ICT (MSIT), Korea, under Grants No. NRF-2022R1A2C1011646, RS-2024-00416036, RS-2025-03392969, and RS-2026-25490114, by the Creation of the Quantum Information Science R\&D Ecosystem program through the NRF funded by the Korean government (MSIT) (Grant No. RS-2023-NR068116), by the Quantum Simulator Development Project for Materials Innovation through the NRF funded by MSIT, Korea (Grant No. RS-2023-NR119931), and by the Brain Pool Program funded by the Ministry of Science and ICT through the NRF (Grant No. RS-2025-25446099). J.H. Kim was supported by the Samsung Science and Technology Foundation (Grant No. SSTF-BA2102-04).

\section*{Author contributions}
M. Kang, Y. Prasad, and N.D. Babu equally contributed to this work.
S. Cheon conceived and supervised the project.
M. Kang constructed the models and analyzed the analytical results.
Y. Prasad calculated the optical response of the system.
N. D. Babu and R. Ghadimi compared the quantum weight and optical signatures.
J. H. Kim presented experimental feasibility and potential implementation of this work.
All authors participated in the writing of the manuscript.

\end{document}